\documentclass[a4paper,11pt,reqno]{amsart}
\usepackage{amsmath,amsfonts,amssymb,enumerate}

\newtheorem{theorem}{Theorem}
\newtheorem{lemma}{Lemma}
\newtheorem{proposition}{Proposition}
\newtheorem{definition}{Definition}

\def\A{\mathcal{A}}

\def\R{\mathbb{R}}
\def\C{\mathbb{C}}

\def\K{\mathbb{K}}
\def\g{\mathfrak{g}}

\def\p{\partial}
\def\proof{\noindent\textit{Proof. }}
\def\qed{$\blacksquare$}
\def\suma{\sum_{\alpha=1}^n}
\def\al{\alpha}
\def\CC{\mathbf{C}}
\def\brhd{\blacktriangleright}

\def\what{\widehat}
\def\X{\mathcal{X}}

\begin{document}

\title{The Weyl realizations of Lie algebras and left--right duality}

\author{Stjepan Meljanac}
\address{Ruder Bo\v{s}kovi\'{c} Institute, Theoretical Physics Division, Bijeni\v{c}ka c. 54, HR 10002 Zagreb, Croatia}
\email{meljanac@irb.hr}

\author{Sa\v{s}a Kre\v{s}i\'{c}--Juri\'{c}}
\address{Faculty of Science, Department of Mathematics, University of Split, Teslina 12, 21000 Split, Croatia}
\email{skresic@pmfst.hr}

\author{Tea Martini\'{c}}
\address{Faculty of Science, Department of Mathematics, University of Split, Teslina 12, 21000 Split, Croatia}
\email{teamar@pmfst.hr}

\date{}

\begin{abstract}
We investigate dual realizations of non--commutative spaces of Lie algebra type in terms of formal power series in the Weyl algebra. To each realization of a
Lie algebra $\g$ we associate a star--product on the symmetric algebra $S(\g)$ and an ordering on the enveloping algebra $U(\g)$. Dual realizations of $\g$
are defined in terms of left--right duality of the star--products on $S(\g)$. It is shown that the dual realizations are related to an extension problem
for $\g$ by shift operators whose action on $U(\g)$ describes left and right shift of the generators of $U(\g)$ in a given monomial. Using properties of the
extended algebra, in the Weyl symmetric ordering we derive closed form expressions for the dual realizations of $\g$ in terms of two generating functions
for the Bernoulli numbers. The theory is illustrated by considering the $\kappa$--deformed space.
\end{abstract}


\maketitle


\section{Introduction}
\label{sec-00}

This paper deals with some aspects of realizations of finite dimensional Lie algebras with emphasis on applications to
non--commutative (NC) spaces. Realizations of Lie algebras by vector fields play a major role in group analysis of differential
equations, such as calculation of symmetry groups and group--invariant solutions \cite{Olver, Ovsiannikov}, group
classification of PDE's \cite{Basarab} and construction of difference schemes for numerical solutions of differential
equations \cite{Bourlioux}. Recently, realizations of Lie algebras have been used extensively in the study of NC spaces and their
deformed symmetries. The study of NC spaces is motivated by physical evidence that the classical concept of point at the
Planck scale $\big(l_P=\sqrt{G\hbar/c^3}\approx 1.62\times 10^{-35}\, m\big)$ is no longer valid due to quantum fluctuations.
Einstein's theory of gravity coupled with Heisenberg's uncertainly principle suggests that space--time coordinates should satisfy
uncertainty relations $\Delta \hat x_\mu \Delta\hat x_\nu \geq l_P^2$ (Refs. \cite{Doplicher-1, Doplicher-2}). One of the possible
approaches towards description of space--time structure at the Planck scale is in the framework of NC geometry. In this approach
one introduces non--commutativity in space--time coordinates via the commutation relations $[\hat x_\mu,\hat
x_\nu]=i\theta_{\mu\nu}(\hat x)$. The anti--symmetric tensor $\theta_{\mu\nu}$ generally depends on $\hat x_\mu$ and a
deformation parameter $h\in \R$, and satisfies the classical limit condition $\lim_{h\to 0}\theta_{\mu\nu}=0$. There is also evidence
from string theory suggesting that space--time coordinates are non-commutative \cite{Seiberg-Witten}. The coordinates $\hat
x_\mu$ can be realized as formal power series in the Weyl algebra semicompleted with respect to the degree of differential operator.
Algebraic relations between space--time coordinates lead to various models of NC spaces such as the Moyal space
\cite{Moyal, Szabo-2}, $\kappa$--deformed space \cite{Lukierski-1, Majid-Ruegg} and generalized $\kappa$--deformed
space \cite{Meljanac-7}. A review of applications of NC spaces in physics can be found in Refs. \cite{Li, Aschieri}.

The present paper deals with realizations of NC spaces of Lie algebra type whose coordinates satisfy the Lie algebra
relations $[\hat x_\mu,\hat x_\nu]=\suma C_{\mu\nu\al} \hat x_\al$. An important example is the $\kappa$--deformed space
defined by $[\hat x_\mu,\hat x_\nu]=i(a_\mu \hat x_\nu - a_\nu \hat x_\mu)$ introduced in Refs.
\cite{Lukierski-1, Majid-Ruegg}. The goal of the present work is to generalize the results on realizations of the
$\kappa$--deformed space to an arbitrary finite dimensional Lie algebra $\g$. We also want to describe some general
features of these results in a realization independent setting. This is done by introducing an associative algebra $\mathcal{H}$
which contains the enveloping algebra $U(\g)$ and studying an action of the generators of $\mathcal{H}$ on $U(\g)$.

The paper is organized as follows. In section 2 we outline the results needed in later sections. For a given Lie
algebra $\g$ with basis $\{X_1,X_2,\ldots, X_n\}$ we study extension of $U(\g)$ to an associative algebra
$\mathcal{H}$ generated by $X_\mu$ and $2n^2$ generators $T_{\mu\nu}$ and $T_{\mu\nu}^{-1}$. We introduce an action of
$T_{\mu\nu}$ and $T_{\mu\nu}^{-1}$ on $U(\g)$ describing the right and left shift of basis elements $X_\mu$ in a monomial $X_1^{\nu_1}
X_2^{\nu_2} \ldots X_n^{\nu_n}\in U(\g)$. The actions of
$T_{\mu\nu}$ and $T_{\mu\nu}^{-1}$ are given by the coproducts $\Delta T_{\mu\nu} = \suma T_{\mu\al} \otimes
T_{\al\nu}$ and $\Delta T_{\mu\nu}^{-1}=\suma T_{\al\nu}^{-1}\otimes T_{\mu\al}^{-1}$.

Section 3 introduces realizations of $\g$ by formal power series of differential operators in a semicompleted Weyl
algebra $\hat \A_n$. To each realization we associate a star--product on the symmetric algebra $S(\g)$ and an ordering prescription on
$U(\g)$. We define left--right dual star--products on $S(\g)$ via $f\, \tilde\star\, g=\tau(f\star g)$ where $\tau$ is the flip operator $\tau (f\ast g)=g\ast f$.
We then study properties of such products in terms of the realizations of $\g$. The role of the generators $T_{\mu\nu}$ is to provide transition
between the left--right dual star--products and also between the associated dual realizations. In section 4 we investigate
in more detail dual realizations of $\g$ in the Weyl symmetric ordering. Using the operators $T_{\mu\nu}$ we find a novel
proof of the dual realizations of $\g$ in terms of the generating functions for the Bernoulli numbers $B_n$
(corresponding to conventions $B_1=\pm 1/2$). In section 5 the theory is illustrated by finding the dual realizations
of the $\kappa$--deformed space and the associated star--products. We show that the star--product is in fact a deformation
quantization of the Lie--Poisson bracket on the dual of the $\kappa$--deformed space.

\section{Extensions of Lie algebras and left--right duality}
\label{sec-01}

Throughout the article $\g$ denotes a Lie algebra of dimension $n$ over a field $\mathbb{K}$ ($\K=\R$ or $\C$). Let
$\{X_\mu\mid 1\leq \mu \leq n\}$ be an ordered basis of $\g$ satisfying the Lie bracket
\begin{equation}\label{01}
[X_\mu,X_\nu] = \suma C_{\mu\nu\al} X_\al.
\end{equation}
The structure constants obey $C_{\mu\nu\al}=-C_{\nu\mu\al}$  and the Jacobi identity
\begin{equation}\label{02}
\sum_{\rho=1}^n \big(C_{\mu\al\rho}\, C_{\rho\beta\nu}+C_{\al\beta\rho}\, C_{\rho\mu\nu}+C_{\beta\mu\rho}\, C_{\rho\al\nu}\big)=0.
\end{equation}
To motivate our discussion let us consider the following simple observation. In
the enveloping algebra $U(\g)$, relations \eqref{01} can be written as
$X_\mu X_\nu = \suma (\delta_{\mu\al} X_\nu +C_{\mu\nu\al}) X_\al$.
In general, if $X=X_1^{\nu_1} X_2^{\nu_2}\ldots X_n^{\nu_n}\in U(\g)$ is a monomial,
then shifting $X_\mu$ to the far right in the product $X_\mu X$ generates polynomials $p_{\mu\al}(X)$ such that
\begin{equation}\label{04}
X_\mu X = \suma p_{\mu\al} (X) X_\al.
\end{equation}
Here $p_{\mu\al}(X)$ is the unique polynomial of the form $p_{\mu\al}(X)=\delta_{\mu\al} X + \textit{lower order terms}$.
Roughly speaking, computation of the polynomials $p_{\mu\al}(X)$ is related to the problem of extending the Lie algebra $\g$ by $n^2$
generators $T_{\mu\nu}$ and defining an action of $T_{\mu\nu}$ on $U(\g)$ such that $T_{\mu\al}\brhd X = p_{\mu\al}(X)$.
Similarly, if $X_\mu$ is shifted to the far left in $X X_\mu$ so that
\begin{equation}\label{05}
X X_\mu = \suma X_\al \tilde p_{\al\mu}(X),
\end{equation}
we want to find another set of generators, say $S_{\mu\al}$, satisfying $S_{\mu\al}\brhd X = \tilde p_{\al\mu}(X)$. The
generators $T_{\mu\nu}$ and $S_{\mu\nu}$ were introduced in construction of the Hopf algebroid structure of the Lie algebra
type NC phase space (see Refs.  \cite{Juric-3, Skoda-2}). In this paper we study properties of $T_{\mu\nu}$ and $S_{\mu\nu}$
associated to a general Lie algebra $\g$. We construct an associative algebra $\mathcal{H}$ by extending $U(\g)$ with
$T_{\mu\nu}$ and $S_{\mu\nu}$, and use this to prove certain results about realizations of $\g$.

Let $\g^L\supset \g$ be the Lie algebra with basis $\{X_\mu, T_{\mu\nu}\mid 1\leq \mu,\nu\leq n\}$ satisfying relations \eqref{01} and
\begin{align}
[T_{\mu\nu},T_{\al\beta}] &= 0,  \label{05-A} \\
[T_{\mu\nu}, X_\lambda] &= \suma C_{\mu\lambda\al} T_{\al\nu}.  \label{05-B}
\end{align}
Our first task is to construct an action of the enveloping algebra $U(\g^L)$ on the subalgebra $U(\g)$.
To keep the notation simple, we identify the elements of $\g^L$ with their canonical images in $U(\g^L)$.

\begin{theorem}\label{tm-01}
Let $\brhd \colon U(\g^L)\otimes U(\g)\to U(\g)$ be a linear map, $a\otimes X\mapsto a\brhd X$, defined by
\begin{equation}\label{07}
1\brhd X=X, \quad X_\mu \brhd X = X_\mu X, \quad T_{\mu\nu}\brhd 1 = \delta_{\mu\nu}
\end{equation}
and $(ab)\brhd X = a\brhd (b\brhd X)$ for $X\in U(\g)$ and $a,b\in U(\g^L)$. Then $\brhd$ is a left action of $U(\g^L)$ on $U(\g)$ satisfying
\begin{equation}\label{08-D}
T_{\mu\nu}\brhd (XY) = \suma (T_{\mu\al}\brhd X) (T_{\al\nu}\brhd Y)
\end{equation}
for all $X,Y\in U(\g)$.
\end{theorem}

\proof First we show that \eqref{08-D} is uniquely fixed by the relations in $U(\g^L)$ and conditions \eqref{07}.
Relation \eqref{05-B} and the normalization condition $T_{\mu\nu}\brhd 1 = \delta_{\mu\nu}$ imply
\begin{equation}\label{08}
T_{\mu\nu}\brhd X_\lambda = \delta_{\mu\nu} X_\lambda + C_{\mu\lambda\nu}.
\end{equation}
Now, for any $Y\in U(\g)$, the identity
$T_{\mu\nu}\brhd (X_\lambda Y)=[T_{\mu\nu},X_\lambda]\brhd Y+X_\lambda (T_{\mu\nu}\brhd Y)$
together with Eqs. \eqref{05-B} and \eqref{08} yields
\begin{equation}
T_{\mu\nu}\brhd (X_\lambda Y) = \suma C_{\mu\lambda\al} (T_{\al\nu}\brhd Y)+X_\lambda (T_{\mu\nu}\brhd Y)
=\suma (T_{\mu\al} \brhd X_\lambda) (T_{\al\nu}\brhd Y).
\end{equation}
This shows that decomposition \eqref{08-D} holds for monomials $X$ of degree one. By induction, assume that \eqref{08-D} holds for monomials $X$ of degree $k$.
Then for $k+1$ degree monomials we have
\begin{multline}
T_{\mu\nu}\brhd \big((X_\lambda X) Y\big) = \suma (T_{\mu\al}\brhd X_\lambda) \big(T_{\al\nu}\brhd (XY)\big) \\
=\sum_{\beta=1}^n \Big(\suma (T_{\mu\al} \brhd X_\lambda)(T_{\al\beta}\brhd X)\Big) (T_{\beta\nu}\brhd Y)
=\sum_{\beta=1}^n \Big(T_{\mu\beta} \brhd (X_\lambda X)\Big) (T_{\beta\nu}\brhd Y).
\end{multline}
By linearly extending the action we find that Eq. \eqref{08-D} holds for all $X,Y\in U(\g)$. In order to prove that the action is well--defined, it suffices to show
that the defining relations of $U(\g)$ are in the kernel of $\brhd$, and that the action is consistent with the relations in $U(\g^L)$. This is obvious
for the action of $X_\mu$, hence we only consider $T_{\mu\nu}$.
Using Eq. \eqref{08-D} we find
\begin{equation}
T_{\mu\nu}\brhd [X_\al, X_\beta] = \delta_{\mu\nu} [X_\al, X_\beta]+ \sum_{\rho=1}^n \big(C_{\mu\al\rho} C_{\rho\beta\nu}+C_{\beta\mu\rho} C_{\rho\al\nu}\big).
\end{equation}
In combination with Eq. \eqref{08}, this implies
\begin{equation}
T_{\mu\nu}\brhd \Big([X_\al,X_\beta]-\sum_{\rho=1}^n C_{\al\beta\rho} X_\rho\Big) = \sum_{\rho=1}^n \big(C_{\mu\al\rho} C_{\rho\beta\nu}+C_{\beta\mu\rho} C_{\rho\al\nu}+
C_{\al\beta\rho} C_{\rho\mu\nu}\big) = 0
\end{equation}
due to the Jacobi identity \eqref{02}. To show consistency of the action with relation \eqref{05-A} note that $[T_{\al\beta},T_{\mu\nu}]\brhd X_\rho = 0$. By induction, assume
that $[T_{\al\beta},T_{\mu\nu}]\brhd X=0$ for all monomials of degree $k$. Then a short computation shows that
\begin{multline}
[T_{\al\beta},T_{\mu\nu}]\brhd (X_\rho X) =\\ X_\rho [T_{\al\beta},T_{\mu\nu}]\brhd X +
\sum_{\kappa=1}^n C_{\al\rho\kappa} [T_{\kappa\beta},T_{\mu\nu}]\brhd X + \sum_{\kappa=1}^n C_{\mu\rho\kappa} [T_{\al\beta},T_{\mu\nu}]\brhd X =0.
\end{multline}
Hence, $[T_{\al\beta},T_{\mu\nu}]\brhd X=0$ for all $X \in U(\g)$. Finally,
consistency with relation \eqref{05-B} follows from Eqs. \eqref{08-D}--\eqref{08} and  noting that for any monomial $X\in U(\g)$ we have
\begin{equation}
[T_{\mu\nu},X_\lambda]\brhd X =\suma (\delta_{\mu\al} X_\lambda +C_{\mu\lambda\al}) (T_{\al\nu}\brhd X) - X_\lambda (T_{\mu\nu}\brhd X)
=\Big(\suma C_{\mu\lambda\al} T_{\al\nu}\Big)\brhd X.
\end{equation}
This completes the proof. \qed

We remark that $T_{\mu\nu}\brhd X$ can be computed recursively from
\begin{equation}
T_{\mu\nu}\brhd (X_\al X) = X_\al \big(T_{\mu\nu}\brhd X\big) + \sum_{\rho=1}^n C_{\mu\al\rho}\, \big(T_{\rho\nu}\brhd X\big).
\end{equation}
The next result shows that the action of $T_{\mu\al}$ on $X$ generates precisely the polynomials $p_{\mu\al}(X)$ defined by Eq. \eqref{04}.

\begin{lemma}\label{lm-01}
Let $X$ be a monomial in $U(\g)\subset U(\g^L)$. If $X_\mu$ is shifted to the far right in the product $X_\mu X$, then
\begin{equation}\label{18}
X_\mu X =\suma (T_{\mu\al}\brhd X)\, X_\al.
\end{equation}
\end{lemma}

\proof For monomials of degree one, Eq. \eqref{18} follows directly from Eq. \eqref{08}.
By induction, assume that \eqref{18} holds for monomials $X$ of degree $k$. Then, in view of Eq. \eqref{08-D} we find
\begin{align}
X_\mu(X_\lambda X) &= \suma (T_{\mu\al}\brhd X_\lambda)\, (X_\al X) =\suma (T_{\mu\al}\brhd X_\lambda) \Big(\sum_{\beta=1}^n T_{\al\beta} \brhd X\Big) X_\beta \notag \\
& = \sum_{\beta=1}^n\Big[\suma (T_{\mu\al}\brhd X_\lambda) (T_{\al\beta}\brhd X)\Big] X_\beta =
 \sum_{\beta=1}^n \Big(T_{\mu\beta}\brhd(X_\lambda X)\Big)X_\beta.
\end{align}
Hence, Eq. \eqref{18} holds for all monomials $X\in U(\g)$. \qed

As noted earlier, shifting $X_\mu$ to the far left
in the product $X X_\mu$ generates polynomials $\tilde p_{\al\mu}(X)$ defined by Eq. \eqref{05}. The following result is analogous to theorem \ref{tm-01}.
It describes an extension of the Lie algebra $\g$ by $n^2$ commuting generators $S_{\mu\al}$  such that
$S_{\mu\al}\brhd X=\tilde p_{\mu\al}(X)$. $S_{\mu\nu}$ can be regarded as element of the formal inverse matrix $[T_{\mu\nu}]^{-1}$,
hence we denote $S_{\mu\nu}=T_{\mu\nu}^{-1}$.

\begin{theorem}
Let $\g^R$ be the Lie algebra with basis $\{X_\mu, T_{\mu\nu}^{-1}\mid 1\leq \mu,\nu \leq n\}$ defined by
relations \eqref{01} and
\begin{equation}\label{23-A}
[T_{\al\beta}^{-1},T_{\mu\nu}^{-1}]=0, \quad [T_{\mu\nu}^{-1},X_\lambda]=\suma C_{\lambda\al\nu}\, T_{\mu\al}^{-1}.
\end{equation}
Then there exists a left action $\brhd \colon U(\g^R)\otimes U(\g)\to U(\g)$ satisfying
\begin{alignat}{2}
&1\brhd X = X, & \qquad   &X_\mu \brhd X = X_\mu X,  \label{22} \\
&T_{\mu\nu}^{-1}\brhd 1 = \delta_{\mu\nu}, & \qquad   &T_{\mu\nu}^{-1}\brhd (XY) = \suma (T_{\al\nu}^{-1}\brhd X) (T_{\mu\al}^{-1}\brhd Y)  \label{23}
\end{alignat}
for all $X,Y\in U(\g)$.
\end{theorem}
The action of $T_{\mu\nu}^{-1}$ is computed recursively from $T_{\mu\nu}^{-1}\brhd X_\lambda = \delta_{\mu\nu}X_\lambda-C_{\mu\lambda\nu}$ and
\begin{equation}
T_{\mu\nu}^{-1}\brhd (X_\al X) = X_\al \big(T_{\mu\nu}^{-1}\brhd X\big)-\sum_{\rho=1}^n C_{\rho\al\nu} \big(T_{\mu\rho}^{-1}\brhd X\big), \quad X\in U(\g).
\end{equation}
Using induction as in lemma \ref{lm-01} one can prove that the right multiplication
by $X_\mu$ can be written as
\begin{equation}\label{28}
X X_\mu = \suma X_\al\, (T_{\mu\al}^{-1}\brhd X).
\end{equation}
At this point it seems natural to extend the enveloping algebra $U(\g)$ by both sets of generators $T_{\mu\nu}$ and $T_{\mu\nu}^{-1}$.
If such extension exists, then the normalization conditions for $T_{\mu\nu}$ and $T_{\mu\nu}^{-1}$ necessarily imply that
$\suma (T_{\mu\al}^{-1} T_{\al\nu})\brhd 1 =\suma (T_{\mu\al} T_{\al\nu}^{-1})\brhd 1 = \delta_{\mu\nu}$. This suggests that $U(\g)$ should have a natural inclusion into
a unital associative algebra defined as follows.
\begin{definition}\label{def-01}
Let $\mathcal{H}$ be a unital associative algebra with generators $X_\mu, T_{\mu\nu}$ and $T^{-1}_{\mu\nu}$, $1\leq \mu,\nu \leq n$,
subject to relations \eqref{01}, \eqref{05-A}--\eqref{05-B}, \eqref{23-A} and the additional relations
\begin{equation}\label{29}
\suma T_{\mu\al}^{-1} T_{\al\nu} = \suma T_{\mu\al} T_{\al\nu}^{-1} = \delta_{\mu\nu}.
\end{equation}
\end{definition}
It is straightforward, albeit lengthy, to verify that $\mathcal{H}$ is well defined, i.e. that the defining relations for $\mathcal{H}$ are consistent.
Clearly, we have the embedding $\g \hookrightarrow \mathcal{H}$ where $\mathcal{H}$ inherits the actions of $U(\g^L)$ and $U(\g^R)$ on the subalgebra $U(\g)$.
In the rest of the paper important role is played by the elements of $\mathcal{H}$ defined by
\begin{equation}\label{32-A}
Y_\mu=\suma X_\al T_{\mu\al}^{-1}.
\end{equation}
The elements $Y_\mu$  generate the left--right dual of the Lie algebra $\g$ introduced in Sec. \ref{sec-02}. We note that
Eq. \eqref{28} implies $Y_\mu \brhd X = \suma X_\al (T_{\mu\al}^{-1}\brhd X) = X X_\mu$.
Hence, $X_\mu$ and $Y_\mu$ act as left and right multiplication operators on $U(\g)$ since $X_\mu\brhd X=X_\mu X$ and $Y_\mu \brhd X = X X_\mu$
for all $X\in U(\g)$. Furthermore, relations \eqref{01} and \eqref{23-A} imply that
\begin{equation}\label{XY}
[X_\mu,Y_\nu]=\suma \big([X_\mu,X_\al] T_{\nu\al}^{-1}-X_\al [T_{\nu\al}^{-1},X_\mu]\big)=0.
\end{equation}
reflecting commutativity of the left and right multiplication in $U(\g)$.

In the following section we consider realizations of $\g$ by formal power series of differential operators. To each realization of $\g$ we
associate a star--product on the symmetric algebra $S(\g)$. Using realizations of the generators $T_{\mu\nu}$ and $T_{\mu\nu}^{-1}$
we construct left--right dual realizations of $\g$ that correspond to left--right dual star--products on $S(\g)$. In the course of our discussion we present a novel
proof of the Weyl symmetric realization of $\g$ found in Ref. \cite{Durov}. The theory presented here is illustrated by means of the $\kappa$--deformed space in section \ref{sec-04}.

\section{Realizations of Lie algebras and left--right duality of associated star--products}
\label{sec-02}

Let $\g_h$ be a Lie algebra over $\K$ defined by
\begin{equation}\label{36}
[X_\mu,X_\nu]=\suma C_{\mu\nu\al}(h)\, X_\al, \quad 1\leq \mu,\nu \leq n.
\end{equation}
We assume that the structure constants $C_{\mu\nu\al}(h)$ depend on a deformation parameter $h\in \R$ suc  h that $\lim_{h\to 0} C_{\mu\nu\al}(h)=0$.
Any Lie algebra $\g$ can be deformed in this way by simply rescaling the structure constants of $\g$, $C_{\mu\nu\al}\mapsto h C_{\mu\nu\al}$.
To simplify the notation, we omit explicit dependence of $C_{\mu\nu\al}$ on $h$.
The enveloping algebra $U(\g_h)$ is the coordinate algebra of the NC space defined by relations \eqref{36}.
In this section we study realizations of $X_\mu$ by formal power series of differential operators in a semicompleted Weyl algebra.
Recall that the $n$--th Weyl algebra $\A_n$ is a unital associative algebra over $\K$ generated by
$x_1,\ldots ,x_n,\p_1,\ldots, \p_n$ satisfying the commutation relations $[x_\mu,x_\nu]=[\p_\mu,\p_\nu]=0, \quad [\p_\mu,x_\nu]=\delta_{\mu\nu}$.
(In the physics literature this is usually called the Heisenberg algebra). The algebra
$\A_n$ has a faithful representation on the vector space of polynomials $\K[x_1,\ldots, x_n]$ where $x_\mu$ stands for multiplication operator by $x_\mu$
and $\p_\mu$ is the partial derivative $\p/\p x_\mu$. We define $\hat \A_n$ to be the semicompletion of $\A_n$ by the order of differential operators.
Thus, $\hat \A_n$ contains formal power series in $\p_\mu$ but only polynomial expressions in $x_\mu$.

\begin{definition}\label{def-02}
A realization of the Lie algebra \eqref{36} is a Lie algebra monomorphism $\varphi\colon \g_h\to \hat \A_n$ defined on the basis of $\g$ by
\begin{equation}\label{32}
\varphi(X_\mu) = \suma x_\al\, \varphi_{\al\mu}(\p),
\end{equation}
where $\varphi_{\al\mu}(\p)$ is a formal power series in $\p_1,\ldots ,\p_n$ depending on $h$ such that\\ $\lim_{h\to 0}\varphi_{\al\mu}(\p)=\delta_{\al\mu}$.
\end{definition}
The map $\varphi$ extends to a unique homomorphism of associative algebras $\varphi \colon U(\g_h)\to \hat \A_n$. The coordinates
$\hat x_\mu = \sum_{\al=1}^n x_\al\, \varphi_{\al\mu}(\p)\in \hat \A_n$ are interpreted as deformations of ordinary coordinates $x_\mu$ since
$\lim_{h\to 0} \hat x_\mu = x_\mu$. Let $\what \X$ and $\X$ denote the subalgebras of $\hat \A_n$ generated by $\hat x_1, \ldots, \hat x_n$ and
$x_1, \ldots, x_n$, respectively. Since $\varphi$ is injective and $[\hat x_\mu, \hat x_\nu]=\suma C_{\mu\nu\al} \, \hat x_\al$, the algebra
$\what \X$ is isomorphic with $U(\g_h)$ (and $\X$ is trivially isomorphic with $S(\g_h)$).
The commutation relations for $\hat x_\mu$ hold if and only if the functions $\varphi_{\mu\nu}$ satisfy a system of formal PDE's:
\begin{equation}
\suma \left(\frac{\p \varphi_{\lambda\mu}}{\p \p_\al}\, \varphi_{\al\nu}-\frac{\p \varphi_{\lambda\nu}}{\p \p_\al}\, \varphi_{\al\mu}\right)=\suma C_{\mu\nu\al}\, \varphi_{\lambda\al}, \quad
1\leq \mu,\nu \leq n.
\end{equation}
This is generally an under--determined system admitting infinitely many solutions parameterized by arbitrary real--analytic functions. The order in which $x_\al$ and $\varphi_{\al\mu}(\p)$
appear in the realization is immaterial since any linear combination
\begin{equation}
\hat x_\mu = c \suma x_\al \varphi_{\al\mu}(\p)+(1-c)\suma \varphi_{\al\mu}(\p)x_\al
\end{equation}
is also a realization of $\g_h$. Hermitian realizations are obtained for $c=1/2$ (see Ref. \cite{Kovacevic-2}). In the rest of the paper we set $c=1$. Examples of different realizations of
NC spaces such as the $\kappa$--deformed space, generalized $\kappa$--deformed space and $su(2)$--type NC space were found in
Refs. \cite{Meljanac-7, Durov, Meljanac-5, Meljanac-4, Skoda, Meljanac-6}.

Realizations of Lie algebras are related to two important concepts: ordering on the enveloping algebra $U(\g_h)\simeq \what \X$ and star--product
on the symmetric algebra $S(\g_h)\simeq \X$. To establish the connection we introduce a left action
$\rhd \colon \A_n \otimes \X\to \X$, $a\otimes f \mapsto a\rhd f$, defined by
\begin{equation}\label{38}
x_\mu \rhd f = x_\mu f, \quad \p_\mu \rhd f = \frac{\p f}{\p x_\mu},  \quad (ab)\rhd f = a\rhd (b\rhd f), \quad f\in \X.
\end{equation}
The action extends to formal power series in $\hat \A_n$ in the obvious way.
For a given realization $\varphi \colon \g_h\to \hat \A_n$ we define the vector space isomorphism
$\Omega_\varphi \colon \what \X\to \X$ by $\Omega_\varphi(\hat f) = \hat f \rhd 1$. Note that for $\hat f=\hat x_{\mu_1} \hat x_{\mu_2}\ldots \hat x_{\mu_k}$ we have
$\Omega_\varphi (\hat f) = x_{\mu_1} x_{\mu_2}\ldots x_{\mu_k}+p_{k-1}$
where $p_{k-1}$ is a polynomial of degree $k-1$ in the variables $x_{\mu_1}, x_{\mu_2}, \ldots, x_{\mu_k}$. The isomorphism $\Omega_\varphi$ induces a
star--product on the algebra $\X$ as follows.

\begin{definition}
The star--product $\star \colon \X\otimes \X \to \X$ associated to realization $\varphi \colon \g_h\to \hat \A_n$ is defined by
\begin{equation}\label{12}
f\star g = \Omega_\varphi \big(\Omega_\varphi^{-1}(f)\, \Omega_\varphi^{-1}(g)\big), \quad f,g\in \X.
\end{equation}
\end{definition}
The algebra $\X^\star = (\X,+,\star)$ is a unital associative algebra which is isomorphic to $\what \X$ since
$\Omega_\varphi (\hat f \hat g) = \Omega_\varphi (\hat f) \star \Omega_\varphi (\hat g)$ for all $\hat f, \hat g\in \what \X$. Furthermore,
the generators of $\X^\star$ satisfy the commutation relations $x_\mu \star x_\nu - x_\nu \star x_\mu = \sum_{\al=1}^n C_{\mu\nu\al} \, x_\al$.
A few remarks about the star--product are in order. The star--product is a deformation of the commutative product in $\X$ since $f\star g = fg+O(h)$.
In deformation quantization, for a given Poisson manifold $(M,\{\, ,\, \})$ one looks for a formal
star--product such that $f\star g - g\star f = ih \{f,g\}$ $(mod \; h^2)$ for $f,g\in C^\infty (M)[[h]]$. The proof of existence of star--products for a
general Poisson manifold was given in Ref. \cite{Kontsevich} by Kontsevich'c formality theorem. In our approach the starting point is not a Poisson manifold but
the non--commutative algebra $\what \X$. The vector space isomorphism $\Omega_\varphi \colon \what \X\to \X$ is then used to transfer the non--commutative
multiplication in $\hat \X$ to the star--product in $\X$. Here the star--product \eqref{12} is treated formally since it may fail to converge when extended to
power series in the variables $x_1,\ldots, x_n$. For technical issues about convergence see Ref. \cite{Skoda}.

Next we introduce left--right duality of the star--product \eqref{12}. Let $\g_h$ be the
Lie algebra with basis $\{Y_1,Y_2,\ldots, Y_n\}$ closing the bracket relations
\begin{equation}\label{48-02}
[Y_\mu,Y_\nu]=-\suma C_{\mu\nu\al} Y_\al, \quad 1\leq \mu,\nu\leq n.
\end{equation}
We say that $\tilde \g_h$ is the ``left--right dual'' of the Lie algebra \eqref{36}.
Although $\g_h$ and $\tilde \g$ are trivially isomorphic via $X_\mu \mapsto -Y_\mu$, the relation between their realizations is generally non--trivial.

\begin{definition}\label{def-04}
The star--products $\star$ and $\tilde \star$ associated with realizations $\varphi\colon \g_h \to \hat \A_n$ and $\tilde \varphi \colon \tilde \g_h \to \hat \A_n$,
respectively, are left--right dual if
\begin{equation}\label{22-1}
f\star g = g\, \tilde \star\, f, \quad f,g\in \X.
\end{equation}
\end{definition}
We remark that the notion of duality introduced here refers to the flip operator $\tau$ since $f\, \tilde \star\, g = \tau (f\star g)$, and is not related to standard duality
between coordinates and momenta in the Weyl algebra. For future reference we refer to $\varphi$ and $\tilde \varphi$ as dual realizations.
It is not difficult to show that the following result characterizes such realizations.

\begin{lemma}\label{lm-03}
Let $\varphi$ and $\tilde \varphi$ be realizations of Lie algebras \eqref{36} and \eqref{48-02} given by
\begin{equation}
\varphi (X_\mu)\equiv \hat x_\mu =\suma x_\al\, \varphi_{\al\mu}(\p), \quad \tilde \varphi (Y_\mu)\equiv \hat y_\mu = \suma x_\al\, \tilde \varphi_{\al\mu}(\p).
\end{equation}
Then the star--products $\star$ and $\tilde \star$ are left--right dual if and only if $[\hat x_\mu,\hat y_\nu]=0$ for all $\mu,\nu=1,\ldots ,n$.
\end{lemma}
The vector space isomorphism $\Omega_\varphi^{-1} \colon \X\to \what \X$ associates to a realization $\varphi$ an ordering on the algebra $\what \X\simeq U(\g_h)$ by mapping the
standard basis of $\X$ to a basis of $\what \X$. Of particular interest is the Weyl symmetric realization which makes $\Omega_{\varphi^{-1}}$ the symmetrization map.
This property is characterized by
\begin{equation}\label{85}
\Big(\sum_{\mu=1}^n k_\mu\, \hat x_\mu\Big)^m \rhd 1 = \Big(\sum_{\mu=1}^n k_\mu x_\mu\Big)^m, \quad \forall k_\mu \in \K, \; m\geq 1.
\end{equation}
In the following section we derive an explicit form of the Weyl symmetric realization and compute its left--right dual. The approach followed here differs from
that in Ref. \cite{Durov} since the key role is played by the algebra $\mathcal{H}$ introduced in definition \ref{def-01}.

\section{The Weyl symmetric realization of the algebra $\mathcal{H}$}
\label{sec-03}

This section deals with the realization of the algebra $\mathcal{H}$ that corresponds to the Weyl symmetric ordering on the
subalgebra $U(\g_h)\subset \mathcal{H}$. The realization uses the generating function for the Bernoulli numbers $B_k$,
\begin{equation}\label{gen-function}
\psi(t)\equiv \frac{t}{1-e^{-t}} = \sum_{k=0}^\infty \frac{(-1)^k}{k!}\, B_k\, t^k
\end{equation}
(with convention $B_1=-1/2$).
Let $\CC$ be the $n\times n$ operator--valued matrix with elements
$\CC_{\mu\nu}=\suma C_{\mu\al\nu}\p_\al$ where $C_{\mu\al\nu}$ are the structure constants of the Lie algebra \eqref{36}.
Let $e^{\CC}$ denote the formal matrix exponential $e^{\CC}=\sum_{k=0}^\infty {\CC}^k/k!$.
\begin{theorem}\label{tm-03}
The algebra $\mathcal{H}$ in definition \ref{def-01} admits the following realization:
\begin{equation}\label{78-A}
\hat x_\mu = \suma x_\al\, \psi_{\mu\al}(\CC), \quad \widehat T_{\mu\nu}=(e^\CC)_{\mu\nu}, \quad \what T^{-1}_{\mu\nu}=(e^{-\CC})_{\mu\nu},
\end{equation}
where $\psi_{\mu\nu}(\CC)$ denotes the $(\mu,\nu)$ element of the matrix $\psi(\CC)$.
\end{theorem}

\proof Define the realization of $T_{\mu\nu}$ by $\widehat T_{\mu\nu}=(e^\CC)_{\mu\nu}$.
Then clearly $[\what T_{\mu\nu},\what T_{\al\beta}]=0$. Next we seek a realization of $X_\mu\in U(\g^L_h)\subset \mathcal{H}$ defined by
$\hat x_\mu = \suma x_\al \varphi_{\al\mu}(\p)$
such that $\what T_{\mu\nu}$ and $\hat x_\lambda$ close relations \eqref{05-B}. It is shown in lemma \ref{lm-04} (see Appendix) that the formal
derivative of $\what T_{\mu\nu}$ is given by
\begin{equation}\label{70}
\frac{\p}{\p_\lambda} \what T_{\mu\nu}= \sum_{\al,\beta=1}^n C_{\mu\al\beta} \left(\frac{1-e^{-\CC}}{\CC}\right)_{\lambda\al}\what T_{\beta\nu},
\end{equation}
hence
\begin{equation}\label{71}
[\what T_{\mu\nu}, \hat x_\lambda] = \sum_{\al,\beta=1}^n C_{\mu\al\beta}\left[\sum_{\kappa=1}^n \left(\frac{1-e^{-\CC}}{\CC}\right)_{\kappa\al}
\varphi_{\kappa\lambda}(\p)\right]\what T_{\beta\nu}.
\end{equation}
If we choose $\varphi_{\kappa\lambda}(\p) = \psi(\CC)_{\lambda\kappa}$, then
\begin{equation}\label{75}
[\what T_{\mu\nu},\hat x_\lambda]=\sum_{\beta=1}^n C_{\mu\lambda\beta}\, \what T_{\beta\nu}.
\end{equation}
For this choice of the realization $\varphi_{\mu\nu}(\p)$, the generators $\hat x_\mu$ are given by the power series
\begin{equation}\label{75-A}
\hat x_\mu = \suma x_\al \, \psi_{\mu\al}(\CC) = \sum_{k=0}^\infty \suma \frac{(-1)^k}{k!}\, B_k\, x_\al (\CC^k)_{\mu\al}.
\end{equation}
Next we show that $[\hat x_\mu,\hat x_\nu]=\suma C_{\mu\nu\al} \, \hat x_\al$. Since the algebra $\hat \A_n$ is associative,
the operators $\hat x_\mu, \what T_{\al\beta}\in \hat \A_n$ satisfy the Jacobi identity
$[[\hat x_\mu,\hat x_\nu], \what T_{\al\beta}]+[[\hat x_\nu, \what T_{\al\beta}],\hat x_\mu]+[[\what T_{\al\beta},\hat x_\mu],\hat x_\nu]=0$.
Substituting \eqref{75} into this identity and using Eq. \eqref{02} we find
\begin{equation}\label{78}
[[\hat x_\mu,\hat x_\nu], \what T_{\al\beta}] = \sum_{\lambda,\rho=1}^n C_{\mu\nu\rho}\, C_{\rho\al\lambda}\, \what T_{\lambda\beta}.
\end{equation}
The matrix $\psi(\CC)$ is invertible, hence $x_\mu = \suma \hat x_\al (\psi(\CC)^{-1})_{\mu\al}$.
This implies that $[\hat x_\mu,\hat x_\nu]=\sum_{\rho=1}^n \hat x_\rho \, \theta_{\mu\nu\rho}(\p)$
for some formal power series $\theta_{\mu\nu\rho}(\p)\in \hat \A_n$. Consequently, Eq. \eqref{75} yields
\begin{equation}\label{79}
[[\hat x_\mu,\hat x_\nu],\what T_{\al\beta}]=\sum_{\rho=1}^n [\hat x_\rho,\what T_{\al\beta}]\, \theta_{\mu\nu\rho}(\p) =
\sum_{\lambda,\rho=1}^n \theta_{\mu\nu\rho}(\p)\, C_{\rho\al\lambda}\, \what T_{\lambda\beta}.
\end{equation}
Comparing equations \eqref{78} and \eqref{79}, and taking into account that the matrix $\what T=e^\CC$ is regular, we find
$\sum_{\rho=1}^n \theta_{\mu\nu\rho}(\p)\, C_{\rho\al\lambda} = \sum_{\rho=1}^n C_{\mu\nu\rho}\, C_{\rho\al\lambda}$.
Thus, the power series $\theta_{\mu\nu\rho}(\p)$ has only the zero--order term $\theta_{\mu\nu\rho}^0\in \K$.
We claim that $\theta_{\mu\nu\rho}^0=C_{\mu\nu\rho}$. Note that the action \eqref{38} yields
$[\hat x_\mu,\hat x_\nu]\rhd 1 = \sum_{\rho=1}^n \theta_{\mu\nu\rho}^0 \, x_\rho$. A short computation using expansion \eqref{75-A} shows that
\begin{equation}
\hat x_\mu \rhd x_\nu = x_\mu\, x_\nu + \frac{1}{2}\sum_{\rho=1}^n C_{\mu\nu\rho}\, x_\rho.
\end{equation}
Therefore,  $[\hat x_\mu, \hat x_\nu]\rhd 1 = \hat x_\mu \rhd x_\nu-\hat x_\nu \rhd x_\mu= \sum_{\rho=1}^n C_{\mu\nu\rho}\, x_\rho$
which proves that $\theta_{\mu\nu\rho}^0 = C_{\mu\nu\rho}$. Thus the generators $\hat x_1,\hat x_2,\ldots, \hat x_n$ close the Lie algebra \eqref{36}.

In a similar fashion one can show that the algebra $U(\g^R_h)\subset \mathcal{H}$ admits the realization given by Eq. \eqref{75-A} and
$\widehat T^{-1}_{\mu\nu}=(e^{-\CC})_{\mu\nu}$. Obviously, $\suma \what T_{\mu\al}\, \widehat T^{-1}_{\al\nu}=\suma \widehat T^{-1}_{\mu\al}\, \what T_{\al\nu}=\delta_{\mu\nu}$,
which proves that \eqref{78-A} defines a realization of $\mathcal{H}$. \qed

We note that the realizations of the generators of $\mathcal{H}$ are defined in terms of the structure constants $C_{\mu\nu\lambda}$ which describe the
adjoint representation of the Lie algebra $\g_h$.
The realization \eqref{75-A} was found in Ref. \cite{Durov} where several different proofs were given using direct computation, formal geometry and a
coalgebra structure. The proof given here is based on the realization of the extended algebra $\mathcal{H}$
which is then utilized to construct the left--right dual realizations of the algebra $\g_h$.
We remark that the star--product associated to realization \eqref{75-A} appears implicitly in Ref. \cite{Gutt} as the Gutt star--product.

\begin{theorem}
The realization of the Lie algebra $\g_h$ given by \eqref{75-A} satisfy the symmetrization property \eqref{85}.
\end{theorem}

\proof We prove relation \eqref{85} by induction on $m$. The claim is obvious for $m=1$ since $\hat x_\mu \rhd 1 = x_\mu$. Assume that
Eq. \eqref{85} holds for some $m>1$. Then by the induction assumption
\begin{equation}\label{86}
\Big(\sum_{\mu=1}^n k_\mu \hat x_\mu\Big)^{m+1}\rhd 1 = \Big(\sum_{\mu=1}^n k_\mu \hat x_\mu\Big)\rhd \Big(\sum_{\mu=1}^n k_\mu x_\mu\Big)^m.
\end{equation}
Define homogeneous polynomials $P_m(x)=\big(\sum_{\mu=1}^n k_\mu\, x_\mu\big)^m$. We write the realization of $\hat x_\mu$ as
$\hat x_\mu = x_\mu + \sum_{k=1}^\infty \suma a_k\, x_\al\, (\CC^k)_{\mu\al}$
where $a_k = (-1)^k\, B_k/k!$. Substituting this into Eq. \eqref{86} we find
\begin{equation}\label{88}
\Big(\sum_{\mu=1}^n k_\mu \hat x_\mu\Big)^{m+1} \rhd 1 =
P_{m+1}(x)+\sum_{k=1}^\infty \suma a_k\, x_\al \Big(\sum_{\mu=1}^n k_\mu (\CC^k)_{\mu\al}\rhd P_m(x)\Big).
\end{equation}
Define coefficients $K_{\mu\al} = \sum_{\rho=1}^n C_{\mu\rho\al} k_\rho$. Then the action of the operator $(\CC^k)_{\mu\al}$ on $P_m(x)$
is found to be
\begin{equation}\label{94}
(\CC^k)_{\mu\al}\rhd P_m(x) = \frac{m!}{(m-k)!} P_{m-k}(x)\, K^k_{\mu\al}, \quad k\geq 1,
\end{equation}
where
\begin{equation}
K_{\mu\al}^1 = K_{\mu\al}, \quad
K_{\mu\al}^{k} =\sum_{\beta_1,\beta_2,\ldots, \beta_{k-1}=1}^n
K_{\mu\beta_{k-1}}K_{\beta_{k-1}\beta_{k-2}}\ldots K_{\beta_1\al}, \quad k\geq 2.
\end{equation}
Note that $\sum_{\mu=1}^n k_\mu K_{\mu\al}=0$ since $C_{\mu\nu\lambda}=-C_{\nu\mu\lambda}$. This yields
\begin{equation}\label{96}
\sum_{\mu=1}^n k_\mu\, K_{\mu\al}^k = 0 \quad \text{for all}\quad k\geq 1.
\end{equation}
Now, substituting Eq. \eqref{94} into \eqref{88} and using Eq. \eqref{96} we find
\begin{equation}
\Big(\sum_{\mu=1}^n k_\mu \hat x_\mu\Big)^{m+1} \rhd 1 =P_{m+1}(x)=\Big(\sum_{\mu=1}^n k_\mu x_\mu\Big)^{m+1}.
\end{equation}
This completes the proof. \qed

\subsection{Left--right duality in the Weyl--symmetric realization}

Recall that in the associative algebra $\mathcal{H}$ the elements $Y_\mu=\suma X_\al T_{\mu\al}^{-1}$ act as right
multiplication operators by $X_\mu$, $Y_\mu \brhd X = X X_\mu$ for $X\in U(\g)$. In view of theorem \ref{tm-03},
the realization of $Y_\mu$ is given by
\begin{equation}\label{ -A}
\hat y_\mu = \suma \hat x_\al\, \what T^{-1}_{\mu\al} =\suma x_\al \left(e^{-\CC}\psi(\CC)\right)_{\mu\al}.
\end{equation}
Interestingly,
\begin{equation}
\tilde \psi(t)\equiv e^{-t} \psi(t)=\sum_{k=0}^\infty \frac{(-1)^k}{k!}\, \bar B_k\, t^k
\end{equation}
is also a generating function for the Bernoulli numbers $\bar B_k$ with convention $\bar B_1=1/2$.
The following result shows that $\hat x_\mu = \suma x_\al\, \psi_{\mu\al}(\CC)$ and $\hat y_\mu = \suma x_\al\, \tilde \psi_{\mu\al}(\CC)$
actually define dual realizations of the Lie algebra $\g_h$.

\begin{theorem}\label{tm-05}
Let $\g_h$ and $\tilde \g_h$ denote the left--right dual Lie algebras \eqref{36} and \eqref{48-02}, respectively. Define linear maps
$\psi \colon \g_h\to \hat \A_n$ and $\tilde \psi\colon \tilde \g_h\to \hat \A_n$ by
$X_\mu \mapsto \hat x_\mu = \suma x_\al\, \psi_{\mu\al}(\CC)$ and $Y_\mu \mapsto \hat y_\mu = \suma x_\al \, \tilde \psi_{\mu\al}(\CC)$
where $\psi(t)=t/(1-e^{-t})$ and $\tilde \psi(t)=t/(e^t-1)$. Then the associated star--products
\begin{equation}\label{93}
f\star g = \Omega_\psi\big(\Omega^{-1}_\psi (f)\, \Omega^{-1}_\psi (g)\big) \quad \text{and} \quad
f\, \tilde \star\, g = \Omega_{\tilde \psi} \big(\Omega^{-1}_{\tilde \psi}(f)\, \Omega^{-1}_{\tilde \psi}(g)\big)
\end{equation}
are left--right dual, i.e. $f\star g = g\, \tilde \star\, f$ for all $f,g\in \X$.
\end{theorem}

\proof

It was already shown in theorem \ref{tm-03} that $\psi$ is a realization of $\g_h$. In order to show that $\tilde \psi$ is a realization of $\tilde\g_h$, i.e.
$[\hat y_\mu,\hat y_\nu]=-\suma C_{\mu\nu\al} \hat y_\al$, we make use of the relations in the algebra $\mathcal{H}$.
Since $\hat y_\mu$ is of the form  $\hat y_\mu = \suma \hat x_\al\, \what T^{-1}_{\mu\al}$, we have $[\hat x_\mu,\hat y_\nu]=0$ in view of Eq. \eqref{XY}.
This implies that $[\hat y_\mu,\hat y_\nu]=\suma \hat x_\al [\what T^{-1}_{\mu\al}, \hat y_\nu]$.
Furthermore, it follows from Eq. \eqref{23-A} that
$[\what T^{-1}_{\mu\al}, \hat y_\nu]=\sum_{\beta,\rho=1}^n C_{\beta\rho\al}\, \what T^{-1}_{\mu\rho}\, \what T^{-1}_{\nu\beta}$, hence
$[\hat y_\mu,\hat y_\nu] = \sum_{\al, \beta,\rho=1}^n C_{\beta\rho\al}\, \hat x_\al\, \what T^{-1}_{\mu\rho} \, \what T^{-1}_{\nu\beta}$. Expressing $\hat x_\al$ as
$\hat x_\al = \sum_{\kappa=1}^n \hat y_\kappa\, \what T_{\al\kappa}$, the last relation can be written as
\begin{equation}\label{YY}
[\hat y_\mu,\hat y_\nu] = \sum_{\kappa=1}^n \hat y_\kappa\, \Big(\sum_{\al,\beta,\rho=1}^n C_{\beta \rho \al}\, \what T_{\al\kappa} \, \what T^{-1}_{\mu\rho}\,
\what T^{-1}_{\nu\beta}\Big).
\end{equation}
According to proposition \ref{prop-04} (see Appendix) that the operators $\what T_{\mu\nu}=(e^\CC)_{\mu\nu}$ satisfy the identity
\begin{equation}
\sum_{\al,\beta,\rho=1}^n C_{\beta\rho\al}\, \what T_{\al\kappa} \what T^{-1}_{\mu\rho} \what T^{-1}_{\nu\beta} = -C_{\mu\nu\kappa},
\end{equation}
hence Eq. \eqref{YY} yields $[\hat y_\mu,\hat y_\nu]=-\sum_{\kappa=1}^n C_{\mu\nu\kappa} \hat y_\kappa$. Now, since $[\hat x_\mu,\hat y_\nu]=0$,
lemma \ref{lm-03} implies that the induced star--products \eqref{93} are left--right dual. \qed

It is important to note that only in the case of the Weyl symmetric ordering, the dual realization is obtained by the simple transformation
$\tilde \psi_{\mu\al}(\CC)=\psi_{\mu\al}(-\CC)$. In arbitrary orderings there is no simple relation between a realization $\varphi_{\mu\al}(\p)$ and
its dual $\tilde \varphi_{\mu\al}(\p)$.

\section{Dual realizations of the $\kappa$--deformed space}
\label{sec-04}

The $\kappa$--deformed space is the enveloping algebra of the Lie algebra
\begin{equation}\label{98}
[X_\mu,X_\nu] = i(a_\mu X_\nu - a_\nu X_\mu), \quad 1\leq \mu,\nu\leq n.
\end{equation}
This algebra appears in the mathematical framework of deformed (doubly) special relativity theories \cite{Kowalski-Glikman-1, Kowalski-Glikman-2} and it has
applications in quantum gravity \cite{Amelino-Camelia-3} and quantum field theory \cite{Daszkiewicz-1, Dimitrijevic}. The deformation parameter $\kappa=1/|a|$
is usually associated with the Planck mass or quantum gravity scale. The Lie algebra \eqref{98} represents deformations of the Euclidean or Minkowski space,
depending on the metric imposed on the underlying commutative space (obtained in the classical limit $\kappa\to \infty$). Realizations of the Lie algebra \eqref{98}
in different orderings have been investigated in Refs. \cite{Meljanac-5, Meljanac-4, Meljanac-6}. Recently, realizations of Lie superalgebras have been
used to construct graded differential algebras on the $\kappa$--Minkowski space in Refs. \cite{Meljanac-1, Meljanac-2, Meljanac-3, Juric-2, Juric-X}.
Left--right dual realizations of the $\kappa$--Minkowski space were found in Ref. \cite{Kovacevic}. Here we restrict our attention to the $\kappa$--Euclidean space,
although the analysis is easily extended to the $\kappa$--Minkowski space.

For future reference, let $\g_\kappa$ denote the Lie algebra \eqref{98}.
The structure constants of $\g_\kappa$ are given by $C_{\mu\nu\lambda}=i(a_\mu \delta_{\nu\lambda}-a_\nu \delta_{\mu\lambda})$.
Let us define the row vectors $a=(a_1,a_2,\ldots, a_n)$ and $\p=(\p_1,\p_2,\ldots, \p_n)$. Then the opearator--valued matrix $\CC_{\mu\nu}=\suma C_{\mu\al\nu}\p_\al$
can be written as $\CC = ia\otimes \p-(ia\cdot \p) I$
where $I$ is the $n\times n$ identity matrix and $a\cdot \p=\suma a_\al \p_\al$. The powers of $\CC$ are given by
\begin{equation}
\CC^k=(-1)^{k-1} A^{k-1} (ia\otimes \p)+(-1)^k A^k I, \quad k\geq 1,
\end{equation}
where $A$ denotes the differential operator $A=ia\cdot \p$. Using the above expression the matrix $\psi(\CC)$, where $\psi (t)$ is the generating function
\eqref{gen-function}, can be found in closed form:
\begin{equation}
\psi(\CC)=\frac{A}{e^A-1}\, I-\frac{1}{A}\Big(\frac{A}{e^A-1}-1\Big)(ia\otimes \p).
\end{equation}
Hence, the Weyl symmetric realization of the $\kappa$--Euclidean space is found to be
\begin{equation}\label{104}
\hat x_\mu =\suma x_\al \psi_{\mu\al}(\CC) = x_\mu \frac{A}{e^A-1}+ia_\mu (x\cdot \p)\Big(\frac{1}{A}-\frac{1}{e^A-1}\Big).
\end{equation}
This realization appears as a special case of an infinite family of covariant realizations found in Ref. \cite{Meljanac-4}
(see also Ref. \cite{Kovacevic}). Note that in the classical limit we find $\lim_{\kappa\to\infty} \hat x_\mu = x_\mu$, as required.
Similarly, one finds
\begin{equation}
\what T_{\mu\nu}=(e^\CC)_{\mu\nu}=e^{-A} \delta_{\mu\nu}-ia_\mu\p_\nu \frac{e^{-A}-1}{A}, \quad
\what T^{-1}_{\mu\nu}=(e^{-\CC})_{\mu\nu}=e^{A} \delta_{\mu\nu}-ia_\mu\p_\nu \frac{e^{A}-1}{A}
\end{equation}
which provides a realization of the associative algebra $\mathcal{H}$ containing the $\kappa$--deformed space \eqref{98}.
According to theorem \ref{tm-05}, the dual realization in the symmetric ordering is simply given by
\begin{equation}\label{102-A}
\hat y_\mu = \suma x_\al \tilde \psi_{\mu\al}(\CC) = x_\mu \frac{A}{1-e^{-A}}+ia_\mu (x\cdot \p)\Big(\frac{1}{A}-\frac{1}{1-e^{-A}}\Big).
\end{equation}
It is shown in Ref. \cite{Meljanac-4} that the star--product associated with the Weyl symmetric realization \eqref{104} can be written in terms of bi--differential
operators as
\begin{equation}\label{107-A}
f\star g = \exp\Big(\suma x_\al (\Delta \p_\al-\Delta_0 \p_\al)\Big)(f,g)
\end{equation}
where
\begin{equation}\label{107-B}
\Delta_0 \p_\al = \overleftarrow{\p_\al}+\overrightarrow{\p_\al} \quad \text{and}\quad
\Delta \p_\al = \overleftarrow{\p_\al}\; \frac{\tilde\psi(\overleftarrow{A}+\overrightarrow{A})}{\tilde \psi(\overleftarrow{A})}+
\overrightarrow{\p_\al}\; \frac{\psi(\overleftarrow{A}+\overrightarrow{A})}{\psi(\overrightarrow{A})}.
\end{equation}
Here, the operators $\overleftarrow{\p_\al}$ and $\overrightarrow{\p_\al}$ act on $f$ and $g$, respectively, while $\overleftarrow{A}$ and $\overrightarrow{A}$ are defined  by
$\overleftarrow{A}=i\suma a_\al \overleftarrow{\p_\al}$ and $\overrightarrow{A}=i\suma a_\al \overrightarrow{\p_\al}$. As usual, the identity operator stands for
pointwise multiplication. The star--product associated with the dual realization \eqref{102-A} is found to be given by interchanging $\psi$ and $\tilde \psi$ in
Eq. \eqref{107-B}. We then have $f\,\star\, g = g\,\tilde \star\, f$, in agreement with theorem \ref{tm-05}.
Expanding the star--product to first order in the deformation parameter $1/\kappa$ we obtain
\begin{equation}
f\star g = fg+\frac{i}{2} \frac{1}{\kappa} \sum_{\al,\beta=1}^n (a_\al^0\, x_\beta-a_\beta^0\, x_\al) (\p_\al f)(\p_\beta g)+O\Big(\frac{1}{\kappa^2}\Big)
\end{equation}
where $a^0=\kappa a\in \R^n$ is a unit vector. Note that the last equation can be written as $f\star g = fg+(i/2\kappa)\{f,g\}+O\big(1/\kappa\big)$
where
\begin{equation}\label{112}
\{f,g\} = \sum_{\al,\beta=1}^n (a^{0}_\al x_\beta-a^0_\beta x_\al) (\p_\al f)(\p_\beta g)
\end{equation}
is the Lie--Poisson bracket on the dual of the Lie algebra $\g_\kappa^0$ having the
structure constants $C^0_{\mu\nu\lambda}=a^0_\mu \delta_{\nu\lambda}-a^0_\nu \delta_{\mu\lambda}$. Thus, the star--product \eqref{107-A}
corresponding to the Weyl symmetric realization of the Lie algebra \eqref{98} is a deformation quantization of the Poisson
manifold $\big((\g_\kappa^0)^\ast,\{\,,\, \}\big)$. More information on quantizing the dual of a Lie algebra is found in Refs.
\cite{Kathotia, Gutt}.

\section*{Acknowledgements}
The work of S.M. has been fully supported by the Croatian Science Foundation under the project IP--2014--09--9582.

\appendix
\section{Identities for the Weyl symmetric realization}
\label{sec-05}

In this appendix we prove some identities used in the proofs of statements in section \ref{sec-03}.
\begin{proposition}\label{prop-01}
The matrix elements $\CC_{\mu\nu}=\suma C_{\mu\al\nu}\p_\al$ satisfy the following identities
\begin{equation}\label{54}
\suma  (\CC^m)_{\mu\al} C_{\al\lambda\nu} = \sum_{\al,\beta=1}^n \left[\sum_{k=0}^m \binom{m}{k} (-1)^k (\CC^k)_{\lambda\al}
(\CC^{m-k})_{\beta\nu}\right] C_{\mu\al\beta}, \quad m\geq 1.
\end{equation}
\end{proposition}

\begin{proposition}\label{prop-02}
\begin{equation}\label{59}
\frac{\p}{\p \p_\lambda} (\CC^m)_{\mu\nu} = \sum_{\al,\beta=1}^n C_{\mu\al\beta}
\left[\sum_{k=1}^m \binom{m}{k} (-1)^{k-1} (\CC^{k-1})_{\lambda\al} (\CC^{m-k})_{\beta\nu}\right], \quad m\geq 1.
\end{equation}
\end{proposition}
The above propositions are easily proved by induction on $m$.

\begin{lemma}\label{lm-04}
\begin{equation}
\frac{\p}{\p_\lambda} (e^\CC)_{\mu\nu} = \sum_{\al,\beta=1}^n C_{\mu\al\beta} \left(\frac{1-e^{-\CC}}{\CC}\right)_{\lambda\al} (e^\CC)_{\beta\nu}.
\end{equation}
\end{lemma}

\proof Using proposition \ref{prop-02} we find
\begin{equation}\label{66}
\frac{\p}{\p_\lambda} (e^\CC)_{\mu\nu}= \sum_{m=1}^\infty \frac{1}{m!} \frac{\p}{\p_\lambda} (e^\CC)_{\mu\nu}=
\sum_{\al,\beta=1}^n C_{\mu\al\beta}\left[ \sum_{m=1}^\infty \sum_{k=1}^m \frac{1}{m!}
\binom{m}{k} (-1)^{k-1} (\CC^{k-1})_{\lambda\al}
(\CC^{m-k})_{\beta\nu}\right].
\end{equation}
The formal power series in Eq. \eqref{66} can be written in closed form using the Cauchy product
\begin{equation}
\sum_{m=1}^\infty \sum_{k=1}^m A_{k-1} B_{m-k}\,
(\CC^{k-1})_{\lambda\al} (\CC^{m-k})_{\beta\nu}=
\left(\sum_{m=0}^\infty A_m\,
(\CC^m)_{\lambda\al}\right)\left(\sum_{m=0}^\infty B_m\,
(\CC^m)_{\beta\nu}\right)
\end{equation}
with $A_k=(-1)^k/(k+1)!$ and $B_k=1/k!$. Then
\begin{multline}
\sum_{m=1}^\infty \sum_{k=1}^n \frac{1}{m!} \binom{m}{k} (-1)^{-1} (\CC^{k-1})_{\lambda\al} (\CC^{m-k})_{\beta\nu} = \\
\left(\sum_{m=0}^\infty \frac{(-1)^m}{(m+1)!}
(\CC^m)_{\lambda\al}\right) \left(\sum_{m=0}^\infty \frac{1}{m!}
(\CC^m)_{\beta\nu}\right)=
\left(\frac{1-e^{-\CC}}{\CC}\right)_{\lambda\al}
(e^\CC)_{\beta\nu},
\end{multline}
hence
\begin{equation}
\frac{\p}{\p \p_\lambda} (e^\CC)_{\mu\nu} = \sum_{\al,\beta=1}^n C_{\mu\al\beta} \left(\frac{1-e^{-\CC}}{\CC}\right)_{\lambda\al} (e^\CC)_{\beta\nu}.
\end{equation}
\qed

\begin{proposition}\label{prop-04}
\begin{equation}\label{A-8}
\sum_{\al,\beta,\rho=1}^n C_{\beta\rho\al}\, (e^\CC)_{\al\kappa} (e^{-\CC})_{\mu\rho} (e^{-\CC})_{\nu\beta} = -C_{\mu\nu\kappa}.
\end{equation}
\end{proposition}

\proof In view of proposition \ref{prop-01} we have
\begin{align}
\suma C_{\al\mu\kappa}\, (e^\CC)_{\beta\al} &= \sum_{m=0}^\infty \frac{1}{m!} \left(\suma (\CC^m)_{\beta\al} C_{\al\mu\kappa}\right) \notag \\
&=\sum_{\al,\rho=1}^n \left[\sum_{m=0}^\infty \frac{1}{m!} \sum_{k=0}^m \binom{m}{k} (-1)^k (\CC^k)_{\mu\rho} (\CC^{m-k})_{\al\kappa}\right] C_{\beta\rho\al}.
\end{align}
It is easily verified that the sum in the brackets is the Cauchy product
\begin{equation}
(e^{-\CC})_{\mu\rho}\, (e^{\CC})_{\al\kappa} = \sum_{m=0}^\infty \frac{1}{m!} \sum_{k=0}^m \binom{m}{k} (-1)^k (\CC^k)_{\mu\rho} \, (\CC^{m-k})_{\al\kappa},
\end{equation}
hence
\begin{equation}\label{A-11}
\suma C_{\al\mu\kappa}\, (e^\CC)_{\beta\al} = \sum_{\al,\rho=1}^n C_{\beta\rho\al} (e^{-\CC})_{\mu\rho}\, (e^\CC)_{\al\kappa}.
\end{equation}
Multiplying Eq. \eqref{A-11} by $(e^{-\CC})_{\nu\beta}$ and summing over $\beta$ we obtain Eq. \eqref{A-8}. \qed


\end{document}